\documentclass[dvips]{article}
\usepackage{graphicx,amssymb,amsmath,times,wrapfig,units,fancyhdr}
\usepackage[margin=10pt,font=small,labelfont=bf,format=hang,indention=-0.9cm]{caption}
\usepackage{icrctc07}

\title{MINOS Observations of Shadowing in the Muon Flux Underground }
\shorttitle{Observation of Shadowing...}	

\authors{E.W. Grashorn$^{1}$,for the MINOS Collaboration$^{2}$}
\shortauthors{E.W. Grashorn et al.} 
\afiliations{$^1$ Univ. of Minnesota School of Physics \& Astronomy, 116
Church St., Minneapolis, MN 55455, USA\\
        $^2$http://www-numi.fnal.gov/collab/collab.ps}

\abstract{A high significance observation of two muon signals, the
shadow of the sun and moon, 
have been seen by the \unit[5.4]{kt} MINOS Far Detector, at a depth of
\unit[2070]{mwe}. 
The distribution of angular separation of muons near the moon was well
described by a Gaussian, which was used to determine the angular
resolution ($0.34^{\circ}\pm0.07^{\circ}$) and
pointing ($0.3^{\circ}\pm0.05^{\circ}$) of the detector.  
}

\begin{document}

\maketitle

\section{Introduction \& Motivation}\label{sec:intro}
The  MINOS Far Detector is a magnetized scintillator and steel
calorimeter, located in the Soudan Mine in 
northern Minnesota, USA, at a depth of 720~m. 
While primary function of the Far Detector is to detect
neutrinos from Fermilab's $\nu_{\mu}$ beam, the great depth and wide
acceptance of the detector combined with the flat overburden of the
Soudan site allow it to serve as a cosmic-ray muon detector as well.
The detector is composed of 486 8~m octagonal planes 2.54~cm thick,
spaced 5.96~cm.  This 5.4~kt detector is 30~m long and has a total
aperture of $\unit[6.94\times10^6]{cm^2~sr}$.  MINOS is the first underground
experiment able to discriminate 
positively charged particles from those that are negatively charged for
the purpose of CPT violation investigations,
but this feature also allows independent study of positively and
negatively charged cosmic rays. 
  
Optical telescopes use a standard catalogue of stars to establish the
resolution and pointing reliability of a new instrument.  This is not
possible for a cosmic ray detector, as there are no cosmic ray sources
available for calibration.  There are two well observed phenomena in the
otherwise isotropic cosmic ray sky, though they are deficits, not
sources.  The sun and the moon provide a means to study the
resolution and pointing of a cosmic ray detector because they  absorb
incident cosmic rays, causing
deficits from their respective location.  These signals allow a
measure of phenomena associated with cosmic ray propagation and
interaction resulting from geomagnetic 
fields, interplanetary magnetic fields, multiple Coulomb scattering,
etc.  These
extra-terrestrial objects have the same $0.5^{\circ}$ diameter as viewed
from Earth, though the sun shadow is more difficult to observe because
is much farther away and has its own magnetic field that
deflects the charged cosmic rays.  The cosmic ray shadow of the moon has
been measured by air shower arrays (CYGNUS
~\cite{Alexandreas:1990wj}, CASA~\cite{Borione:1993xq} , Tibet~\cite{Amenomori:1993iv} ), as well
as underground detectors (Soudan~2~\cite{Cobb:1999mi},
MACRO~\cite{Ambrosio:1998wv,Ambrosio:2003mz}, L3+C~\cite{Achard:2005az}).  
\section{Data}\label{sec:data}
 This analysis encompassed events recorded over 1339 days, from 1~August~2003 - 31~March~2007, for a total of
1194 live-days, and includes 51.41 million cosmic ray induced muon
tracks.  Cosmic~ray muons were triggered by recording hits on 4/5 planes or
exceeding a
pulse-height threshold and were written to a temporary disk at Soudan
and later sent to Fermilab for reconstruction.  Several cuts were
required to ensure only well reconstructed tracks were included in the
analysis.  
 Pre-analysis and run cuts include: failure of
demultiplexing figure of merit and data taking quality
\cite{Mufson:2007}.  Analysis cuts include: track length less than
2~m, 
number of planes less than 20, $\chi^2_{reco}/ndf>1.0$, and either track vertex
or end point outside of the fiducial volume of the detector.   A total of
30.52 million events survived these cuts for the combined sample. 

The background for this analysis was calculated using a simple
\textsl{Monte Carlo} simulation that exploits two key features of the muons induced by cosmic ray
primaries: the time between consecutive cosmic
ray arrivals follows
a well known distribution \cite{Grashorn:2007} and the cosmic ray sky is
isotropic, 
Thus, a bootstrap method that independently
chooses the arrival time and location in space efficiently simulates a cosmic
ray muon.  This simulation
chose a muon out of the known distribution of
events in the detector (in horizon coordinates), paired it with a random
time chosen from the 
known time distribution, and found the muon's location in celestial
coordinates.  This was done for every muon to create
one background sample; 500 background samples were simulated for a high
statistics background distribution.  
\section{Combined Muon Calibration}\label{sub:combined}
 A further cut, excluding
 tracks with $p<\unit[30]{GeV/c}$, was required to reject muons that were
 greatly affected by energy dependent processes.  The final
 data set included 20.17 million muons. 
To find the one dimensional space angle separation from the moon
and the sun for each muon track, 
 a list of the celestial body's location in
celestial coordinates in one hour increments was obtained from the JPL
HORIZONS \cite{horizons} ephemeris database for May 1, 2003
until May 1, 2007.  The ephemeris data was referenced to the
location of the detector 
in latitude, longitude,
and distance below sea level.   A function was written that 
interpolated the
celestial body's location at the time a particular muon is seen in the
detector.  Then the  
particular muon's arrival direction was compared to the celestial body's
location at that time.    
\begin{figure*}
\begin{center}
\vspace{-10pt}
\begin{minipage}[l]{0.49\linewidth}
\includegraphics[width=0.99\textwidth]{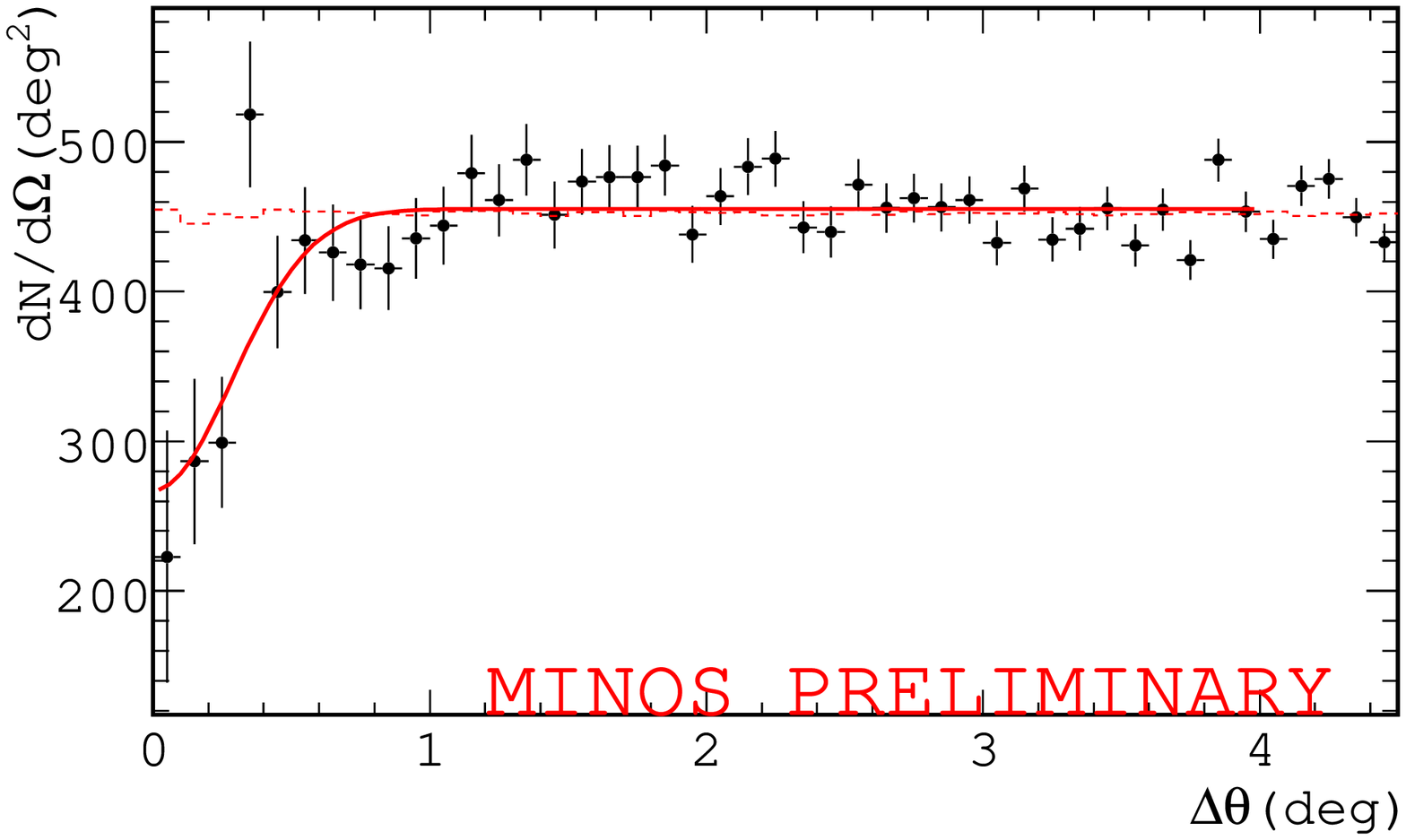}
\end{minipage}
\begin{minipage}[r]{0.49\linewidth}
\includegraphics[width=0.99\textwidth]{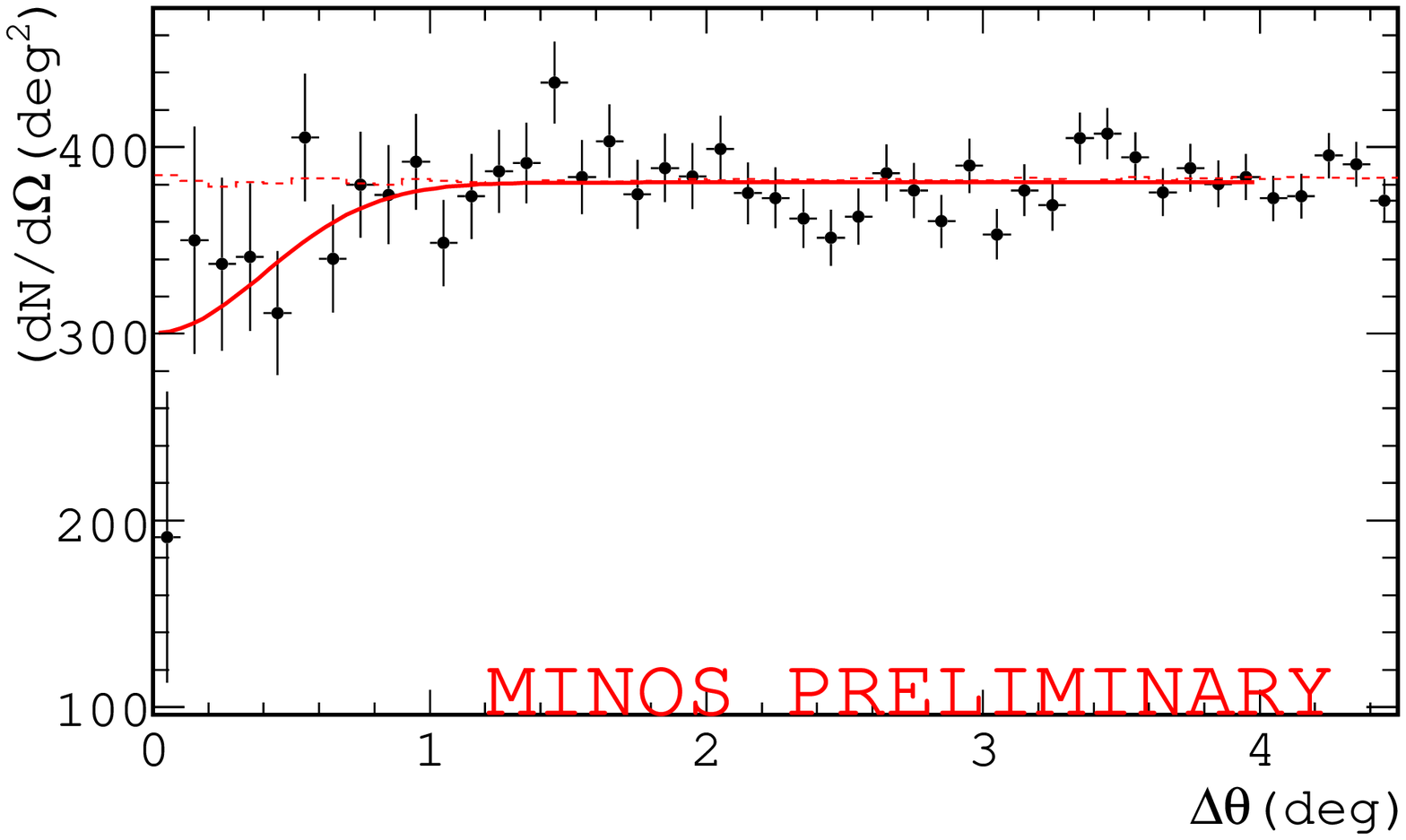}
\end{minipage}
\vspace{-10pt}
\caption{\label{fig:msData} The differential muon flux with respect to
displacement from the moon's (l), and sun's (r) location, 
binned in $0.1{^o}$.  The dashed curve is the calculated background,
while the solid curve is the best fit from eq.~\ref{eq:fit}.} 
\vspace{-20pt}
\end{center}
\end{figure*}
The one
dimensional space angle separation was found using 
the Haversine formula, published in Sky and Telescope
\cite{Sinnott:1984}.  
The muons were binned in $S_{bin} = 0.10^{\circ}$ increments, and
since radial distance from the center of the celestial body is measured over
a two dimensional projection,
the bin solid angle of bin (i) increases when moving out from the center
as $\Delta \Omega_i = (2i-1)*S_{bin}\pi$.  Weighting the number of events in each
bin by the reciprocal of the area resulted in the distribution
$N_i/\Delta \Omega_i$, the differential muon density. 
The $\Delta \theta$ distribution is shown in Fig.~\ref{fig:msData}~(l); from the sun,
Fig.~\ref{fig:msData}~(r), with statistical error bars displayed.  There is a very clear deviation from a flat
distribution in both plots as $\theta \rightarrow 0$, and that deviation is attributed
to muons blocked by the moon and sun, respectively.  The background was
calculated using the method described in Sec. \ref{sec:data}.  As
expected, the backgrounds are nearly flat, with a fit value of $\chi^2/ndf
= 0.12/39$ for the moon and  $\chi^2/ndf= 0.27/39$ for the sun .  This
\textsl{Monte Carlo} is consistent with the premise of 
a sourceless cosmic ray sky.
The significance of
the shadow can be found by fitting to the distribution a function of the form
\cite{Cobb:1999mi}: 
\begin{equation}\label{eq:fit}
\frac{\Delta N_{\mu}}{\Delta \Omega} =
\lambda(1-(R_m^2/\sigma^2)e^{-\theta^2/2\sigma^2})
\end{equation}
where $\lambda$ is the average differential muon flux, $\sigma$ accounts
for smearing from detector resolution, multiple coulomb scattering and
geomagnetic deflection, and $R_m = 0.26^{\circ}$, the angular radius of
the moon.  A fit to eq.~\ref{eq:fit} yields $\chi^2/ndf = 37.9/38$, an
improvement of 16.4 over the liner fit ($\chi^2/ndf = 54.3/39$), with
parameters $\lambda = 483.9 \pm 3.1$ and $\sigma = 0.34^{\circ} \pm
0.07^{\circ}$.  The change in $\chi^2$ over 38 degrees of freedom for
the moon
corresponds to
a $10^-4$ chance probability.
The improvement of $\chi^2$
probability for the sun corresponds to a $10^{-3}$ chance probability.
These results are summarized in Table~\ref{tab:sigtable}.  

Since the tracks of dimuon events, muons that are induced by the same
cosmic ray primary, are nearly parallel 
when they are created, they can be used to find the MINOS
Point Spread Function.
To quantify the absolute pointing of the Far Detector, the MINOS Point
Spread Function will be used to find the two
dimensional contours of the most significant muon deficit caused by the
moon.  This 
analysis 
is not complete at
the time of this writing, however, so in its stead a simple
approximation was used on the one dimensional moon shadow to place an
quantify the pointing of the detector.  A significant shadow
pointing to the apparent location of the moon was found using a one dimensional
space angle separation, a convolution of $\Delta RA, \Delta Dec$. Since
the moon's radius is $0.26^{\circ}$, we can approximate the absolute
pointing of the detector as $0.3^{\circ} \pm 0.05^{\circ}$, with the
error given by one half the bin width in theta.
This analysis assumes that
the pointing is reliable to begin with, so more precise analysis is still to be
performed.  
\section{Charge-Separated Analysis}
\begin{figure*}
 \begin{center}
  \begin{minipage}[l]{0.49\linewidth}
   \includegraphics[width=0.99\textwidth]{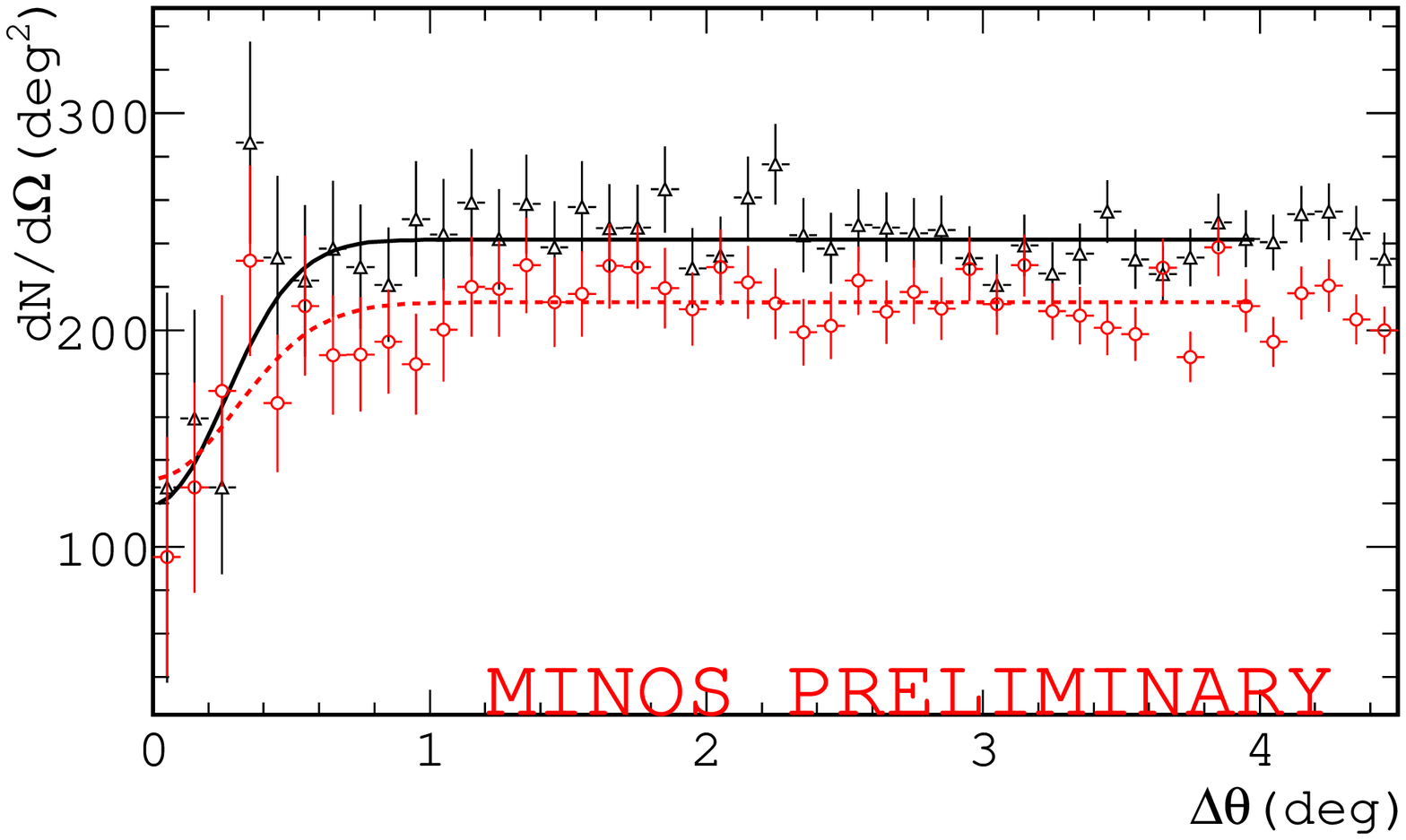}
  \end{minipage}
  \begin{minipage}[r]{0.49\linewidth}
   \includegraphics[width=.99\textwidth]{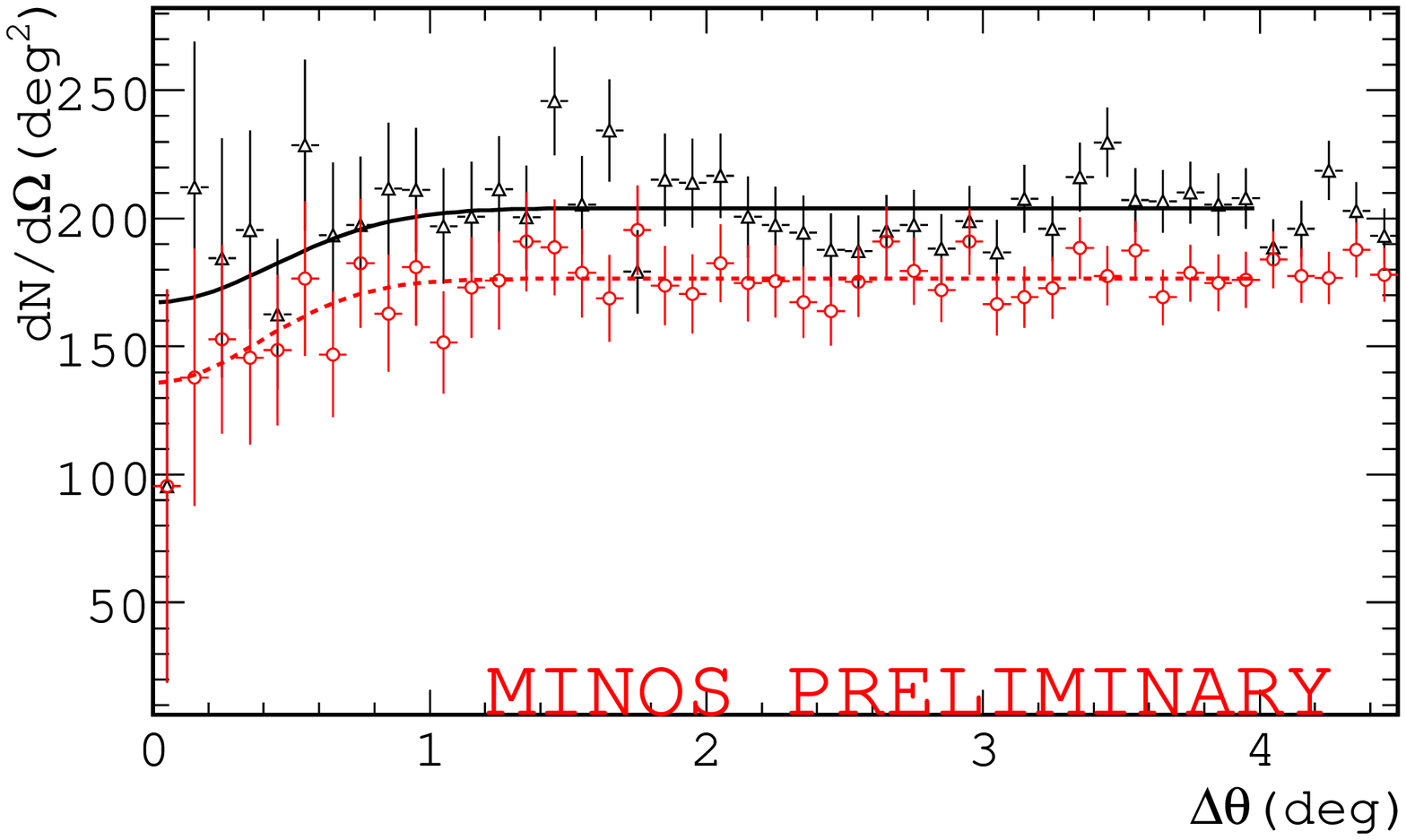}
  \end{minipage}
\vspace{-10pt}
  \caption{\label{fig:mssep} The differential muon flux with respect to
displacement from the moon's location (l), and sun's location (r), 
binned in 0.1${^o}$ increments for $\mu^+$ (open circles) and $\mu^-$,
(open triangles).  The solid curve is the best fit from
eq.~\ref{eq:fit} for $\mu^+$; the dashed curve is for $\mu^-$.  The fit
results are in Table~\ref{tab:sigtable}.} 
\vspace{-10pt}
 \end{center}
\end{figure*}
 For the
charge separated sample, a further cut was required to exclude events
with low confidence charge sign determination.  The curvature of the
track is used to determine the momentum and charge of the particle, so
this cut was charge over momentum divided by the error in the
determination of charge over momentum ($\frac{q/p}{\sigma_{q/p}}>2.2$)~\cite{Adamson:2007vt}.  There
are 2.7 million positively charged and 1.9 million negatively charged
muon in this sample. 
An
analysis similar to what is described in Sec.~\ref{sub:combined} was
performed on the charge separated muon sample.  The charge separated
moon shadow plots can be seen
in Fig.~\ref{fig:mssep}~(l), and the sun shadow plots are shown in
Fig.~\ref{fig:mssep}~(r).  The Gaussian hypothesis gives very little
improvement  over the flat hypothesis, suggesting that the shadow of the
moon or sun has not been seen for the charge separated sample.  
The strong cut $\frac{q/p}{\sigma_{q/p}}$ reduces the statistics by
over four times, from 20 to 4.6 million, eliminating an observable
deficit in the direction of either sun or moon.  This cut does give a
charge ratio of 1.3 for both sun and moon distribution, which is
consistent with the published MINOS result.  This analysis has provided
a flat distribution in the region approaching the moon, so given more
statistics a significant shadow should be observed.


 \begin{table*}
   \centering
  \begin{tabular}{||l|c|c|c|c||} \hline\hline
  \textbf{Distribution}  & \textbf{$\Delta \chi^2$} &
   \textbf{prob.} & \textbf{$\lambda$} & \textbf{$\sigma$} \\\hline \hline
 moon-total &  54.3-37.9 = 16.4 & $ 10^{-4}$ & $483.9 \pm 3.1$ & $0.34 \pm 0.07$
 \\\hline    
 moon-$\mu^+$ & 25.4 - 25.3= 0.1 & N/A  & $56.7 \pm 1.4$ & $ N/A $
 \\\hline   
 moon-$\mu^-$ & 17.4-16.0 = 1.4 & N/A & $43.2 \pm 1.2$ & $ N/A $
 \\\hline   
 sun-total & 48.5 - 40.3 = 8.2 & $10^{-3}$ & $397.4 \pm 2.8$ & $0.399 \pm 0.09$
 \\\hline    
 sun-$\mu^+$ & 23.23 -23.23 = 0  & N/A & $48.2 \pm 19.72$ & $N/A$
 \\\hline   
 sun-$\mu^-$ &  26.68 - 26.68 = 0& N/A & $36 \pm 1.3$ & $N/A$
 \\\hline \hline   
   \end{tabular}
  \caption { Significance for the shadowing observed in each distribution,
 with $\Delta \chi^2 \equiv \chi_{line}^2 - \chi_{gaus}^2$
 \label{tab:sigtable}  }
\end{table*}
\section{Conclusions}
Using  20.17 million muons accumulated over 1194 live-days, the
MINOS Far Detector has observed the cosmic ray shadow of the moon with a
high significance.  Despite the inherent fragility of the
 one-dimensional moon shadow measurement (there were only nine events in
 the bin nearest to the moon), the null hypothesis of the muon deficit
 in the area near to the apparent location of the moon has a probability
 of $10^-4$.
The cosmic ray shadow of the sun over the
 same time period has a chance probability of $10^{-3}$
The shadow of the moon was used to
 approximate both the effective angular resolution of the detector,
$0.34^{\circ}\pm0.07^{\circ}$, and the absolute pointing of the
 detector, $0.30^{\circ}\pm0.05^{\circ}$.  
In agreement with the expectation, no
significant difference was found in the shadowing effects for either
population, save for the charge ratio.
\section{Acknowledgments}
This work was supported by the U.S. Department of Energy and the
University of Minnesota.  Special thanks to the mine crew in Soudan for
their tireless effort keeps the detector up and running.  
\bibliography{ShadowsPaper}
\end{document}